\documentstyle[psfig]{mn2e}
\def\etal{{\it et al.~}}

\def\inventh  	{\textstyle {1 \over \varrho} \displaystyle}
\def\half  	{\textstyle {1 \over 2} \displaystyle}

\def\fourth  	{\textstyle {1 \over 4} \displaystyle}
\def\fifth  	{\textstyle {1 \over 5} \displaystyle}
\def\inv16  	{\textstyle {1 \over 16} \displaystyle}
\def\3ov4  	{\textstyle {3 \over 4} \displaystyle}
\def\4ov3  	{\textstyle {4 \over 3} \displaystyle}
\def\8ov11  	{\textstyle {8 \over 11} \displaystyle}
\def\15ov16  	{\textstyle {15 \over 16} \displaystyle}

\def\beq{\begin{equation}} \def\eeq{\end{equation}}
\def\bea{\begin{eqnarray}} \def\eea{\end{eqnarray}}

\def\AM{ 	\hat{\rm A} } 	
\def\qB{ 	\hat{q}_{_{\hat{\bf B}}} } 	
\def\hchi{ 	\hat{\chi} }

\def\rd{{\rm d}}
\def\sA{{\cal A}}
\def\sC{{\cal C}}
\def\sD{{\cal D}}
\def\Lie{{\cal L}}
 
\def\cN{{\cal N}}
\def\cF{{\cal F}}
\def\cH{{\cal H}}
\def\cO{{\cal O}}

\def\Valf{{\it v}_{_{\rm Alf}}}
\def\Alf{{\rm Alfv\'en}~}

\def\AG{{\rm Araya-G\'ochez\,}}

\def\Prg{p_{\rm r _+ g}}
\def\Prad{p_{\rm rad}}
\def\Pgas{p_{\rm gas}}

\def\PB{p_{_{\rm \bf B}}}
\def\PBphi{p_{_{\rm \bf B_\varphi}}}

 	\def\bk{ {\bf k}}	
\def\bB{ {\bf B}} 	 	
\def\bv{{\bf v}} 	 	\def\bJ{ {\bf J}} 	
\def\bV{{\bf V}} 	
\def\tB{\tilde{B}} 
\def\bOm { {\mbox{\boldmath $\Omega$}} }
\def\bxi { {\mbox{\boldmath $\xi$}} }
\def\unvc{ {\mbox{\boldmath $1$}} }

\def\Del { {\mbox{\boldmath $\nabla$}\!} }
\def\del { {\partial}}
\def\vdel { \vec{\partial}}

\def\UdD{ U \!\! \cdot \!\! \nabla }
\def\udD{ u \! \cdot \!\! \nabla }
\def\dxi{ \dot{\xi} }
\def\dxidD{ \dot{\xi} \! \cdot \!\! \nabla }

\def\hom{\hat{\sigma}}

\def \ltaprx {\lower .1ex\hbox{\rlap{\raise .6ex\hbox{\hskip .3ex
	{\ifmmode{\scriptscriptstyle <}\else 
		{$\scriptscriptstyle <$}\fi}}}
	\kern -.4ex{\ifmmode{\scriptscriptstyle \sim}\else 
		{$\scriptscriptstyle\sim$}\fi}}}
\def\gtaprx {\lower .1ex\hbox{\rlap{\raise .6ex\hbox{\hskip .3ex
	{\ifmmode{\scriptscriptstyle >}\else 
		{$\scriptscriptstyle >$}\fi}}}
	\kern -.4ex{\ifmmode{\scriptscriptstyle \sim}\else 
		{$\scriptscriptstyle\sim$}\fi}}}

\begin{document}


\title{Hydromagnetic Stability \\of a Slim Disk in a Stationary Geometry}

\author [R. A. Araya-G\'ochez]
	{Rafael Angel 
	Araya-G\'ochez\footnote{arayag@tapir.caltech.edu}	\\
  	Theoretical Astrophysics MS 130-33 			\\
  	California Institute of Technology, Pasadena CA 91125	
	} 

\onecolumn  	
\date{}
\pubyear{2002}
\maketitle

\begin{abstract} 
 	The magnetorotational instability originates from the 
 elastic coupling of fluid elements in orbit around a gravitational well.
 Since inertial accelerations play a fundamental dynamical role in the process,
 one may expect substantial modifications by strong gravity 
 in the case of accretion on to a black hole.
 In this paper, we develop a fully covariant, Lagrangian displacement 
 vector field formalism with the aim of addressing these issues 
 for a disk embedded in a stationary geometry with negligible radial flow.
 This construction enables a transparent connection between particle dynamics 
 and the ensuing dispersion relation for MHD wave modes.
 The MRI--in its incompressible variant--is found to
 operate virtually unabated down to the marginally stable orbit; 
 the putative inner boundary of standard accretion disk theory.
 To  get a qualitative feel for the dynamical evolution of the flow
 below $r_{\rm ms}$, we assume a  
 mildly advective accretion flow such that the angular velocity profile
 departs slowly from circular geodesic flow.
 This exercise suggests the turbulent eddies will occur at spatial scales
 approaching the radial distance
 while tracking the surfaces of null angular velocity gradients.   
 The implied field topology, namely large-scale horizontal field domains, 
 should yield strong mass segregation at the displacement nodes 
 of the non-linear modes
 when radiation stress dominates the local disk structure
 (an expectation supported by quasi-linear arguments and 
 by the non-linear behavior of the MRI in a non-relativistic setting). 
 Under this circumstance, baryon-poor flux in horizontal field 
 domains will be subject to radial buoyancy and to the Parker instability, 
 thereby promoting the growth of poloidal field.
\end{abstract}    

\begin{keywords}
MHD---instabilities---black hole physics---gravitational waves
\end{keywords}

\section{Introduction}
 \label{sec:Intro}

	The process of accretion onto compact objects has been long 
 recognized as the primary mechanism to power 
 the most luminous events in space.
 In the traditional picture 
 of Shakura \& Sunyaev (1973) and Novikov \& Thorne (1973),
 entropy is generated and radiated locally from the free energy 
 available in a shear flow with a Keplerian angular velocity profile.
 Two salient oversimplifications of this framework have been the focus 
 of intense research and progress in the last decade: 
 energy advection by the flow and 
 free-energy tapping and angular momentum transport 
 through magnetohydrodynamical processes.  

	The magnetorotational instability 
 or MRI (Velikov 1959, Chandrasekhar 1961, Balbus \& Hawley 1991), 
 justifies the long-sought mechanism for efficient,
 turbulent transport of angular momentum that enables accretion disks 
 to operate with astrophysically interesting mass accretion 
 rates (Pringle 1981).  		
 The importance of this process cannot be overstated:
 By catalyzing accretion into gravitational wells, 
 the MRI enables a plethora of astrophysical phenomena to occur, 
 from protostar formation inside molecular clouds to jet launching in quasars.
 The MRI also holds the key to understand the extraction 
 of free energy from the differential shear flow of otherwise 
 hydrodynamically stable disks (Balbus \etal 1999, Godon \& Livio 2000).

	On the observational front, 
 the wealth of high-quality data from spectral and timing devices aboard 
 space-borne high-energy observatories 
 has turned out the most compelling evidence yet of accretion onto black holes.
 The discoveries of pairs of high frequency quasi-periodic oscillations
 in RXTE X-ray timing data from microquasars
 GRO J1655-40 and GRO 1915+105 (Strohmayer 2001a,b) 
 have brought the spotlight to hydrodynamical models of 
 adiabatic global excitations of the inner disk, 
 a.k.a. diskoseismology models (Perez \etal 1997)
 or relativistic precession models (Stella \etal 1999).
 Interestingly, the role of magnetic fields has been largely ignored
 in spite of clear evidence that QPO's, 
 being non-thermal, hard X-ray phenomena, 
 likely do not originate in the accretion disk proper but rather on a 
 magnetically active accretion disk corona (R. Blandford, Priv.Comm.).
 Likewise, the recent report (Wilms \etal 2001) of the detection of a very 
 broad Fe K$\alpha$ feature on the XMM-EPIC spectrum of MCG-6-30-15,
 has made a very strong case for the inadequacy of standard models
 of energy deposition in accretion disks.  
 The proposed exits to this 
 paradox--extraction of black hole spin energy
 (Blandford \& Znajek 1977), or non-zero torque at the 
 marginally stable circular orbit radius, $r_{\rm ms}$
 (Agol \& Krolik 2000)-- 
 both rely on magnetic coupling between a standard disk
 and the flow inside $r_{\rm ms}$.
 
	On a more exotic front, 
 theoretical progress in our understanding of accretion processes 
 at the most extreme imaginable  
 conditions--stellar-mass black holes hyper-accreting at twelve orders
 of magnitude above the Eddington limit--requires 
 attention to a detailed physical account of highly relativistic 
 accretion flows. 
 Aside from the potential to explain gamma-ray burst phenomenology,
 such studies are chiefly relevant to assess the likelihood of 
 ``failed" supernovae as gravitational wave sources (Fryer \etal 2001).  
 Indeed,
 when neutrino trapping occurs at
 $\dot{M} \gtaprx 1 M_\odot$ sec$^{-1}$ (Popham \etal 1998),
 the associated dynamical stress will mimic 
 the effects of radiation stress in standard disks
 where clumpy accretion ensues (Turner \etal 2001).  
 If the mass fraction in the clumps is large, 
 prolific gravitational waves will be emitted from the mass quadrupole 
 moment associated with the bulk motion of large mass over-densities. 
 Such a scenario will also lead to excitations of the black hole's geometry 
 which, at high values of the spin parameter $a$, can produce
 highly characteristic, monochromatic black hole ringing 
 as the geometry settles towards a quiescent Kerr state
 (\AG 2003). 
 Remarkably, the expectation of a large mass fraction in the clumps 
 is reasonable and justifiable by the physical picture 
 of near-hole accretion presented herein. 

	An outstanding issue yet to be addressed in light of recent
 theoretical progress is our view of black hole accretion inside 
 the marginally stable orbit,
 the putative inner boundary of standard accretion disk theory.
 In particular, very little is concretely known about the inertial 
 effects of strong gravity on the relevant MHD processes.
 Previous work has either assumed pure hydrodynamical flow
 (and energetically negligible energy release)
 or, alternatively, laminar flow under ideal MHD conditions
 (Krolik 1999, Gammie 1999).
 Krolik (1999) has brought out an interesting point:
 Under mere flux freezing conditions the assumption of ballistic orbits
 in the plunging region is {\it never} self-consistent; 
 when the radial velocity component is significant, the magnetic field energy
 density becomes comparable to the rest mass energy density of the matter.

 	In this paper, we address the issue of stability of the magnetic field 
 (co-moving frame) in a stationary, axially symmetric background geometry. 
 Curiously, the two key developments in accretion disk theory of 
 the last few years may have come of age to properly address 
 the problem at hand:
 Inside $r_{\rm ms}$ the accretion flow will be mildly advective,
 with a slightly sub-Keplerian angular velocity profile and possibly 
 supported in part by the radial pressure gradient of a hot MHD fluid 
 with significant relativistic enthalpy 
 (see Popham \& Gammie (1998) solutions 
 for moderate values of $\alpha$ and advected fraction $f$).
 In this spirit, we argue 
 in \S \ref{sec:EndNot}
 that the natural evolution of the MRI inside 
 the marginally stable orbit is at least consistent with this view. 

	The (magnetohydro)dynamics of black hole accretion comprises 
 two important aspects that have received relatively little attention: 
 the effects of radiation pressure 
 (see, however, Blaes \& Socrates and Turner \etal 2001),
 and the effects of strong gravity
 (see footnote 7 of Gammie \& Popham 1998).  
 We will address the former problem in a future paper
 (\AG \,\& Vishniac 2002) 
 while concentrating on general relativity in this one.
 As a background, \S 2 
 looks at the Lagrangian displacement vector field formulation of the MRI 
 concentrating on inertial and compressibility effects.
 In section 3, we develop a fully covariant theory of the instability.
 The intention is to build a theoretical framework from first principles
 in order to avoid missing any subtleties associated to the full
 incorporation of gravitational effects 
 (e.g., reference is made to the Cowling approximation 
 and to the fixing of the gauge associated with the component 
 of the Lagrangian diplacement along the fluid's four-velocity).
 The elastic reponse of the field is computed by noting that the surface 
 of invariance of the Faraday tensor attributes mathematically identical 
 variational properties to the two four-vectors that span it:  
 the magnetic field four-vector and the fluid's four-velocity.
 We then make the minimal modifications to the relativistic fluid equations 
 that allow for the inclusion of a coherent magnetic field
 and undertake a local stability analysis of this field
 in the medium of a slim disk around a rotating black hole,  
 while suppressing compression. 
 The role of compressibility in a photon gas is then briefly assessed.


\section{A Lagrangian Formulation of the MRI in Compressible Media}
 \label{sec:LagForm}

	The MRI is essentially a local instability.
 In the frame of the fluid, the interplay of inertial ``forces" 
 with the elastic coupling of fluid elements creates an unstable situation 
 to the redistribution of specific angular momentum, $\ell$.
 Without the elastic coupling provided by the {\it bending} of field lines, 
 such inertial forces--namely, the shear(tide) and the coriolis terms--induce
 radial epicyclic motions while preserving specific angular momentum 
 in collisionless fluids (e.g., stars in the Galaxy).
 This is related to 
 the Rayleigh criterium for stability of a differentially rotating fluid: 
 $r^{-3}$ d$_r \ell^2 = \kappa^2 \geq 0$, where $\kappa$ 
 is the frequency of epicyclic motions. 

	In the weak field limit, 
 one may construct a dispersion relation 
 quite independently of the specific magnetic field topology: 
 Highly sub-thermal fields, $\Valf/c_s \ll 1$, guarantee that  
 the instability is truly 
local\footnote{
 When the field is non-negligible, $k_\parallel^{-1}$ may approach 
 the disk's pressure scale height and in the case of supra-thermal 
 toroidal fields,
 non-axisymmetric modes have fastest growing wave numbers that may approach 
 the inverse radial scale-length (Foglizzo \& Tagger 1995).
}, occurring at large values of 
 $k_\parallel \simeq \Omega/\Valf ~~(\equiv \bk \cdot \unvc_{\hat{\bB}})$.  
 In this simplified approach,  
 the global disk structure is ignored
 (no curvature nor radial structure) and the response of the
 field amounts to nothing more than providing a restoring force 
 to displacements from equilibrium (Balbus \& Hawley 1992). 
 Indeed, 
 in the {\it horizontal regime of Lagrangian displacement}
 two orthogonal field topologies yield nearly identical 
 mathematical dispersion relations for wavemodes:
 axisymmetric perturbations of a meridional field
 and non-axisymmetric perturbations of a toroidal field. 
 The former case corresponds to the ``classical" Balbus-Hawley instability
 and it's physical relevance is free of controversy. 
 The relative importance of the latter analyses is a more subtle issue.

 	A somewhat technical point--well discussed in the review by 
 Balbus \& Hawley (1998)--is the non-locality induced by shear on wave-modes 
 with Eulerian coordinate phase-dependencies.  For $k_\varphi \neq 0$ modes,
 shear evolves the radial component of wavenumbers according to 
 $k_r(t) = k_{0r} - [{\rm d}_{\ln r} \Omega ] \, k_\varphi t$,
 which means that modes that could be ``unstable" are only so, transiently.
 The maximum instantaneous growth rate occurs when $k_r \rightarrow 0$
 and matches that of the local axisymmetric modes.
 In the end, this issue turns out to be more academic than practical
 but it stresses the importance of treating the instability locally, 
 in co-moving coordinates.  The down side is that   
 this greatly complicates global approaches that rely on eigenmode 
 solutions in Eulerian coordinates extrinsic to the fluid.
 On the other hand, in a local approach azimuthal wavenumbers
 are no longer discrete (Ogilvie \& Pringle 1996)
 and consequently, neither are the co-moving frequencies (see below).

 	A related issue concerns the relevance of non-axisymmetric 
 mode analyses when the magnetic field is not purely toroidal.
 Balbus \& Hawley (1998) argue that the strict ordering of 
 wavenumber components (and narrow phase space) necessitated
 to achieve fastest growth: $k_\varphi \ll k_r \ll k_\theta$,
 ensue in violent poloidal Alfv\'enic couplings that promptly 
 take over the dynamics. 
 Non-axisymmetric modes, however, are important for
 at least two key reasons: 
 {\bf a-} the ordering is not so restrictive when the fields are not weak
 (as needed to explain $\alpha$ values of a few tenths), and
 {\bf b-} {\it compressive},
 non-axisymmetric modes are fundamental to examine energy deposition when
 radiation stress becomes significant (\AG~ \& Vishniac 2002).
 Moreover, because the dispersion relations relate simply
 (at least in the horizontal regime of fastest growth),
 it is rather lucrative to examine both cases at once.  

	Aiming to formulate a fully covariant relativistic theory of the MRI
 in \S \ref{sec:RelEff}, this section conducts the same in three dimensions.
 The linear stability analysis is carried out in terms of the 
 Lagrangian displacement vector field, $\bxi$.  Foglizzo (1995)
 has stressed the usefulness of this approach to account for the 
 polarization of compressive MHD modes.
 A simple meridional stratification profile sets 
 the physical scale-length of the problem: 
 d$_z \ln \rho = \cH^{-1}$, 
 with gas, radiation (and possibly magnetic) pressures
 tracking the unperturbed density
 profile $\rho \cH \Omega = \Prg + \PB$.
 The problem naturally splits in two parts: 
 computation of the inertial/geometric terms 
 \S \ref{subs:GlobAnal}, 
 and computation of the body forces from 
 gas, radiation and electromagnetic stresses
 \S \ref{subs:Compr}.
 We avoid going into the rotating frame from the onset 
 in order to preserve a transparent connection to a 
 ``universal" standard of rest frame
 (to be associated with Boyer-Lindquist coordinates).

\subsection{inertial terms}
 \label{subs:GlobAnal} 

 	Inertial accelerations are 
 geometrically imprint in the connection terms for the covariant
 derivatives of the Eulerian velocity components. 
 For spherical {\it coordinate motion}
 $(\dot{r},\dot{\varphi},\dot{\theta}) 
 \longrightarrow (V^r, V^\varphi, V^\theta)$,
 the only non-trivial connections are 
 $\Gamma^r_{\varphi\varphi} \wedge \Gamma^\varphi_{r\varphi}$. 
 Denoting the Lagrangian time derivative by
 $\rd_t \equiv \del_t + \bV \cdot \Del$,
 the three components of Euler's equation read
\bea
 \rd_t V^r = 
	\del_t V^r + V^j V^r_{,j} + (-r) V^\varphi V^\varphi 
 		&=& g^{r j} {\bf f}_j			\cr
&&\cr
 \rd_t V^\varphi = 
	\del_t V^\varphi + V^j V^\varphi_{,j} + (2/r) V^r V^\varphi 
	 	&=& g^{\varphi j} {\bf f}_j		\cr
&&\cr {\rm and}~~~
 \rd_t V^\theta =
	\del_t V^\theta+ V^j V^\theta_{,j}  
	 	&=& g^{\theta j} {\bf f}_j		\cr
&&\cr
{\rm where~~~~~ {\bf f}} \equiv -{1\over \rho} {\Del p}
	+ {1 \over {4\pi\rho}} \bJ \times \bB, &&
\label{eq:EulCooMot} \eea
 $g^{ij}$ is the flat-space metric for spherical coordinates,
 and $\Del$ will stand for the covariant derivative hereon.

	Assuming an equilibrium from purely azimuthal 
 (but differential) bulk motion
 \bV = $\Omega \unvc_\varphi$, an Eulerian perturbation of such a state 
 $(V^r, V^\varphi, V^z) \longrightarrow (v^r, \Omega + v^\varphi, v^z)$,
 leads to the usual equations associated with a rotating frame and its 
 coriolis and centrifugal terms.  For coordinate motion, Euler equations
 for the perturbations of the fluid read 
\bea
	(\del_t + \Omega \del_\varphi) v^r - 2 r \Omega v^\varphi 
	 	&=& g^{r j} \delta {\bf f}_j 	\cr
&&\cr
	(\del_t + \Omega \del_\varphi) v^\varphi 
	+ (2 \, {\Omega \over r} + \Omega_{,r}) v^r 
	 	&=& g^{\varphi j} \delta {\bf f}_j 	\cr
&&\cr
	(\del_t + \Omega \del_\varphi) v^\theta 
	 	&=& g^{\theta j} \delta {\bf f}_j
\eea
 where $\delta {\bf f}$ stands for the Eulerian perturbation of
 the sum of specific body forces.  The standard form of these equations,
 e.g., for {\it non-coordinate motion} (see, Chandrasekhar 1961),
 may be obtained from Eqs [\ref{eq:EulCooMot}] above 
 by ``dimensionalizing" $V^\varphi$,
 (i.e., in the second Eq, multiplying by $r$ and completing the 
 differential while recalling that the covariant derivative and 
 the metric commute $[\Del,g_{ij}] = \emptyset$).

 	Next, one switches dynamical variables from the Euler
 velocity perturbation, $\bv$, to the Lagrangian displacement, $\bxi$, 
 using the first order relation between Lagrangian and Eulerian variations,
 $\tilde{\Delta} = \delta + \bxi \cdot \Del$,  
whilst denoting\footnote{
 The tilde indicates that this form of Lagrangian 
 displacement--which is generally non-unique--has had its gauge ``fixed"
 in accordance to the non-relativistic regime. 
 Mathematically, this amounts to a choice of Universal time direction,
 $\unvc_t$ (e.g., unaffected by the fluid's motion), 
 while adopting the gauge fixing condition 
 $\xi \cdot \unvc_t \doteq \emptyset$.
} 
 $\tilde{\Delta} \bV \equiv {\rm d}_t \bxi$ and $\delta \bV \equiv \bv$
 (see, e.g., Chandrasekhar \& Lebovitz 1964, Lynden-Bell \& Ostriker 1967)
\beq
	\bv = \{\del_t + \bV \cdot \Del\} \, \bxi - (\bxi \cdot \Del) \bV
	\rightarrow \, i \sigma \bxi - \xi^r \Omega_{,r} \unvc_\varphi.
\label{eq:DynSwi} \eeq
 The algebraic relation follows from the assumption of differential 
 rotation and from writing exp$\;i(\omega t + m \varphi + k_z z)$ 
 dependencies for $\bxi$. 
 Note that the connection coefficients in Eq [\ref{eq:DynSwi}]
 cancel one another and that 
 $\sigma \doteq \omega + m\Omega$ 
 denotes the co-moving frequency of the perturbations. 

	These geometrical equations have their more traditional equivalents 
 in the so-called shearing sheet approximation where a co-moving,
  ``locally Cartesian frame" 
 $(\hat{r}, \hat{\varphi}, \hat{\theta}) \rightarrow (x,y,z)$,
 is used along with the linearized shear velocity field, 
 $\bV(x) = [{\rm d}_{\ln r} \Omega] \, x \unvc_y$,
 to treat the problem locally while introducing the coriolis terms by hand.
 Defining the Cartesian derivative operator $\vdel$, then
 the equivalent to Eq [\ref{eq:DynSwi}] is
 $(\del_t + \bV \cdot \vdel) \bxi = \bv + \bxi \cdot \vdel \, \bV$,
 which has Galilean invariance in the sense that  
 Lagrangian time derivatives produce co-moving frequency 
 factors in the dispersion relation: 
 $(\del_t + \bV \cdot \vdel)\bxi \equiv d_t \bxi \rightarrow i \sigma \bxi$.

	The equations of motion (EoM)
 for the (coordinate) Lagrangian displacement are 
\bea
 	(- \sigma^2 + 2 \Omega \, r\Omega_{,r}) \xi^r 
 	- 2 r\Omega \, i \sigma \, \xi^\varphi  
	&=& \delta {\bf f}^r		\cr
&&\cr
 	- \sigma^2 \xi^\varphi 
 	+ 2 \, {\Omega\over r} \, i \sigma \, \xi^r  
	&=& \delta {\bf f}^\varphi  	\cr
&&\cr
	- \sigma^2 \xi^\theta &=& \delta {\bf f}^\theta.
\eea
 Note that the Eulerian shear term, $\propto \unvc_\varphi$,
 becomes the tide term, $\propto \unvc_r$, in terms of $\bxi$.

 	Let us re-cast these equations in a more compact form
\beq
	\ddot{\xi}^i 
	+ 2 \Gamma^i_{jk} V^j \dot{\xi}^k
	- 2 \Gamma^i_{jk} V^j (v - \dot{\xi})^k
	= \delta {\bf f}^i
\label{eq:TheEssence} \eeq
 where each over-dot stands for a factor of $i \sigma$
 (from a Lagrangian time derivative). 
 In the shearing sheet approximation, these equations correspond 
 to the Hill equations for {\it non-coordinate} motion 
 (Chandrasekhar 1961, Balbus \& Hawley 1992)
 $\ddot{\bxi} + 2 \, \bOm \times \dot{\bxi}  
	+ 2 r \Omega \Omega_{,r} \, \xi_x \, \unvc_x = \delta {\bf f}$.

\subsection{compressibility} 
 \label{subs:Compr}

 The Lagrangian perturbation of mass density and the Eulerian perturbation
 of the field follow from mass and magnetic flux conservations 
 (recall the non-relativistic relation 
 $\tilde{\Delta} = \delta + \bxi \cdot \Del$)  
\beq
	{\Delta \rho \over \rho} = - \Del \cdot \bxi, ~~~~~~
	\delta \bB = \Del \times (\bxi \times \bB).
\eeq
 The latter equation includes possible gradients of the background field
 $\del \bB \neq \emptyset$; however, in the spirit of examining the
 instability as a local phenomenon 
 the global structure of the field is ignored herein. 
 The Lorentz force variation (co-moving frame) may then be written as 
\bea
	\delta \left( {1 \over {4\pi\rho}} \bJ \times \bB \right)
 	&=& \Valf^2 \,\, \times \,\,
 	\left\{	\,\,  \Del (\Del \cdot \bxi) 	
		+ \nabla^2_{\hat{\bB}} \bxi 	
		- \nabla_{\hat{\bB}} \Del (\unvc_{\hat{\bB}} \cdot \bxi)
		- \unvc_{\hat{\bB}} \nabla_{\hat{\bB}} (\Del \cdot \bxi) 
		\right\} \cr
&&\cr
&& 	\stackrel{\nabla \rightarrow i\bk}{\longrightarrow} 
	-\Valf^2 \times
	\{	(k_i \xi_i - k_{\hat{\bB}} \xi_{\hat{\bB}}) \bk
	+	 k^2_{\hat{\bB}} \bxi 	
	- 	\unvc_{\hat{\bB}} k_{\hat{\bB}} k_i \xi_i
	\} 	
\label{eq:Loren_F} \eea 
 where the scalar operator 
 $\nabla_{\hat{\bB}} \equiv  \unvc_{\hat{\bB}} \cdot \Del$,
 and $\unvc_{\hat{\bB}}$ is a unit vector in the direction of 
 the unperturbed field.  Note that the term 
 $(k_i \xi_i - k_{\hat{\bB}} \xi_{\hat{\bB}}) \doteq k_\perp \xi_\perp$
 may be interpreted as a restoring force due to the 
 {\it compression} of field lines 
 (distinct from line bending, Foglizzo \& Tagger 1995).
  
	The Lagrangian variation of the specific pressure gradient 
 contains two terms (Lynden-Bell \& Ostriker 1967): 
 one $\propto \Delta \rho^{-1}$ and another $\propto \Delta \Del \Prg$.
 In terms of the displacement vector, the first term is  
 proportional to the equilibrium value of $\Del \Prg$ 
 which is 
negligible\footnote{
 the variation of the mass density is also negligible
 when the focus is on the effects of radiative heat conduction:
 Loss of pressure support out of compressive modes involves 
 only the pressure term (\AG \& Vishniac 2002).
} in the local treatment (proportional to a radial gradient). 
 For the same reason, the Eulerian and Lagrangian variations 
 of the pressure ``force" are identical.

 The thermodynamic pressure term is then given by
\beq
 	- \delta \left( {1 \over \rho} \Del \Prg \right) 
	= \Gamma \, {\Prg \over \rho} \Del (\Del \cdot \bxi) 
	\, \stackrel{\nabla \rightarrow i\bk}{\longrightarrow} \,
	c_s^2 \, \bk \, k_i \xi_i,
\label{eq:Therm_F} \eeq
 where, for heterogeneous media, 
 $\Gamma \equiv d_{[\ln \rho]} \ln p$ 
 represents a generalized adiabatic index 
 (see, e.g., Chandrasekhar 1939, Mihalas \& Mihalas 1984). 

	Putting the above equations together, one gets
 the EoM in Fourier-space 
\beq
	\ddot{\xi}^i 
	+ 2 \Gamma^i_{jk} V^j \dot{\xi}^k
	- 2 \Gamma^i_{jk} V^j (v - \dot{\xi})^k
	=
	g^{ij} [ \{ 	c_s^2 			\, \bk 
	+ \Valf^2 ( \bk - k_{\hat{\bB}}	\, \unvc_{\hat{\bB}}) \} 
	(\bk \cdot \bxi) 
	+ \Valf^2 (	k^2_{\hat{\bB}} \, \bxi
	- k_{\hat{\bB}} \xi_{\hat{\bB}} \, \bk) ]_j.
\label{eq:k-EoM} \eeq
 which agrees with the matrix de-composition of 
 Foglizzo \& Tagger (1995) 
 in the case of a purely toroidal field embedded in
 a gas with adiabatic index $\Gamma = 1$.

	Reckoning of fluid compressibility has complicated somewhat  
 the equations of motion.  Yet, these generally unwieldy equations 
 simplify greatly in the regime of fastest growth
 (a.k.a. the horizontal regime)
 and for two ideal field topologies of interest.
 When the field is meridional, 
 the fastest growth modes have $\xi_{\hat{\bB}} \doteq \emptyset$,
 and $\bk \simeq k_{\hat{\bB}} \unvc_{\hat{\bB}}$
 thus yielding a simple isotropic elastic response
 $\propto - \Valf^2 (ik_{\hat{\bB}})^2 \, \bxi$.

 	Alternatively, when the field is purely toroidal, 
 the meridional component of Eq [\ref{eq:k-EoM}] yields
 an anisotropy constraint:
 $\Valf^2 \, (k_\perp \xi_\perp) = -c_s^2 \, (\bk \cdot \bxi)$
 (Foglizzo \& Tagger 1995),
 which allows for a straightforward solution in this regime. 
  Defining $\Lambda$ through
\beq
	1-\Lambda = -\frac {\Del \cdot \bxi} {ik_\parallel \xi_\parallel}	
	= {2 \Theta \over {\Gamma + 2 \Theta}} 
\label{eq:Compress} \eeq
 where $\Theta \equiv \PBphi/\Prg$, 	
 the dispersion relation out of Eq [\ref{eq:k-EoM}]
 reads 
\beq 		\hom^4 
	- \{ (	\Lambda + 1) \qB^2 + \hchi^2 \} \, \hom^2
	+ 	\Lambda \, \qB^2 \{  \qB^2 +  4 \AM \}
	= \emptyset,
\label{eq:disper} \eeq
 where all frequencies  
 are normalized to the rotation rate, 
 $\AM \equiv \half {\rm d}_{\ln r} \ln \Omega$ is the Oort A ``constant",
 $\hchi^2 \equiv 4(1+\AM)$ is the squared of the epicyclic frequency, and 
 $\qB \equiv ({\bk} \cdot {\bf v}_{\rm Alf}) / \Omega$ is a 
 frequency related to the component of
 the wave vector along the field (in velocity units). 

	The non-axisymmetric modes of fastest growth conform with 
 (\AG \& Vishniac 2002)
\beq
	\qB^2 =	- 2 \AM 
	+ ( {\textstyle {{1 + \Lambda}\over{2\Lambda}} \displaystyle} ) 	
	\times	  \left\{ -{2 \Lambda \AM^2 \over \sD} \right\},
~~~~~~~{\rm where}~~~~~~~ 
	\sD \equiv 1 
		+  ({\textstyle { {1 - \Lambda}\over2 }\displaystyle}) \AM	
		+  \sqrt{1 + (1 - \Lambda) \AM} ,	
\label{eq:qB-om} \eeq
 while the expression in the curly brackets identifies with
 the negative root of the dispersion relation.

 	Note that the compressibility of non-axisymmetric modes is 
 imprint on the deviations of $\Lambda$ from unity.
 From Eq [\ref{eq:Compress}] one reads that the degree of 
 compression of these modes gets stronger with the (toroidal) field 
 strength and, naturally, with a softer equation of state.
 In the companion paper (\AG \& Vishniac 2002), 
 we find that when radiation pressure begins to dominate the disk dynamics, 
 an ``ultra soft" effective (adiabatic) index 
 accentuates the effects of mode compressibility. 
 On the other hand, 
 setting $\Lambda \doteq 1$ and re-orienting the field vertically  
 produces the standard (incompressible) dispersion form for the 
 Balbus-Hawley instability of a meridional field in the horizontal regime.

\section{General Relativistic Effects in the Cowling Limit}
 \label{sec:RelEff} 

	In contrast to the Newtonian case,
 the formulation of a covariant theory of accretion disk oscillations
 requires more than mere application of the Lagrangian rate of change
 operator d$_t$ (or its relativistic counterpart d$_\tau$) 
 to the Eulerian velocity perturbation.  
 This is insufficient to carry out a normal mode analysis because 
 of the freedom associated with the choice of coordinates. 
 It is much more useful and proper to 
 free the eigenmodes from the coordinate representation;
 treating them rather as intrinsic to the physical system.  
 It is here that a Lagrangian construction comes in handy.

 	A working covariant definition of the 
 Lagrangian displacement is that of  
 a vector field that moves a fluid element's world line 
 from its unperturbed position  in spacetime to its perturbed one.
 The fundamental relation between the
 Lagrangian (following the fluid's world line) and 
 Eulerian (taken at a fixed coordinate point)
 variational operators is
\beq 	\Delta = \delta + \Lie_\xi
\label{eq:RelLag_v_Eul} \eeq
 where $\Lie$ stands for the Lie derivative. 
 
	An elemental use of this relation involves 
 particle number conservation (Schutz \& Sorkin 1977):
 with the use of a number flux density
 $\cN^\nu \equiv n \, \sqrt{-g} \, U^\nu$, 
 such a law reads $\Delta \cN \doteq \emptyset$,
 where $g = \det|g^{\mu\nu}|$ and $U^\nu$ is the
 four-velocity of the fluid.  	
 In the Cowling approximation, $\delta g \doteq \emptyset$,
 and in the absence of co-moving sources (or sinks) of particles, 
 one gets for the variations
 of the four-velocity of an ideal fluid:
\beq
 	\Delta U \doteq 0 = \delta U + \Lie_\xi U 
\label{eq:U_Var}\eeq 
 which demonstrates that Eulerian perturbations of the four velocity,
 $\delta U \equiv u$, obey $u^\nu = -\Lie_\xi U^\nu$.

	The connection to the Newtonian limit is recuperated upon 
 identifying $c \, U \cdot \Del$ 
 with the convective (or material) rate of change
 $c \, U \cdot \Del \stackrel{c \rightarrow \infty }{\longrightarrow} 
 (\del_t + \bV \cdot \Del)$ 
 so that, with $\tilde{\Delta} \bV \equiv {\rm d}_t \bxi$, one has
\beq
 	\tilde{\Delta} = \delta + \bxi \cdot \Del
\label{eq:NewLag_v_Eul} \eeq
 (see, e.g., Chandrasekhar 1964, Lynden-Bell \& Ostriker 1967, and 
 compare $\delta \bV \equiv \bv$ with Eq [\ref{eq:DynSwi}]).

	The fluid particles that constitute a thin accretion disk 
 (with negligible radial inflow) embedded in 
 a Kerr spacetime geometry have unperturbed four velocity 
 $U^\nu = \gamma (\unvc_t + \Omega \unvc_\varphi)$
 where $\unvc_t = (1,0,0,0)$ and $\unvc_\varphi = (0,0,1,0)$ 
 are the Killing vector fields of the stationary, axisymmetric geometry
 and where
 $\gamma$ is the ``redshift" factor of the fluid elements at fixed radius
 $\gamma = U^t = {\rm d}_\tau t$. 	
 In terms of the Lagrangian displacement vector field,
 each Lie derivative with respect to one of the Killing vector fields
 of the geometry ``brings down" a wavenumber co-factor in the dispersion 
 relation (modulo spatial gradients of the four velocity).
 This leads to an algebraic relation between $\xi$
 and $u \equiv \delta U$ in the case of a differentially rotating fluid:
\bea
	u 	&=& \Lie_{\gamma (\unvc_t + \Omega \unvc_\varphi)} \, \xi  \cr
&&\cr
		&=& \dot{\xi}
		 -  \gamma \, \xi^r \, \Omega_{,r} \, \unvc_\varphi
		 -  U \, \xi \cdot \nabla \ln \gamma.
\label{eq:Eul2Lag_D} \eea
 Here $\dot{\xi} \equiv  i \sigma \gamma \xi^\nu$, with 
 $\sigma = \omega + m \Omega$ the co-moving frequency of the perturbation
 as measured at asymptotic infinity (see Ipser \& Lindblom 1992).

	The relativistic generalization of Eq [\ref{eq:DynSwi}] is found 
 upon projecting $\xi$ on the 3-surface perpendicular to the (unperturbed)
 four velocity. 
 With $h^{\alpha\beta} \equiv U^\alpha U^\beta + g^{\alpha\beta}$
 the projection operator, one has 
\beq
 	h^\alpha_{~\beta} u^\beta 	\equiv 
	\hat{u}^\alpha 		= i \sigma \gamma \hat{\xi}^\alpha
	- \gamma \, \xi^r \, \Omega_{,r} \, \hat{\unvc}^\alpha_\varphi
\label{eq:RelDynSwi} \eeq
 while fixing the gauge freedom associated with 
 the component of the Lagrangian displacement 
 perpendicular to space-like hypersurfaces
 (Schutz \& Sorkin 1977), i.e. along the local ``time" direction.  
 Note that the requirement of unit normal for the {\it perturbed} velocity,
 $U + u$, under the Cowling approximation fixes the gauge accordingly:
 $2 U^\alpha u_\alpha = 
 - \, U^\alpha U^\beta \delta g_{\alpha \beta} \doteq 0$. 


	Dynamical conservation laws for an ideal fluid in the presence 
 of a large-scale electromagnetic field are written succinctly through the
 Einstein-Maxwell equation (Cowling approximation)
\[	T^{\mu\nu}_{~~;\nu} - F^{\mu\nu} J_\nu  = 0,
\]
 where the first term stands for the matter stress and the second 
 equals the Maxwell stress.  The notation is standard fare:
 $F$ is the electromagnetic field tensor
 and $J = n\, e\, U$ 
 is the four-current.
 Ideal MHD makes things easy by stating that the electric field 
 in the co-moving frame vanishes everywhere.  
 Since the latter is the contraction of the field tensor 
 with the four-velocity, it follows that $F_{\mu\nu} J^\nu  = 0$, and 
 the four-acceleration from the Maxwell stress vanishes as well
 (but not its perturbation).

 	We shall concern ourselves with the material stress first.

 	Denoting the relativistic enthalpy of the fluid by 
 $\varrho \equiv \rho + \varepsilon + p$,
 it is straightforward to show that 
 for a non-dissipative, ideal fluid such that 
 $T^{\mu\nu} = \varrho U^\mu U^\nu + p g^{\mu\nu}$,
\beq 	
	T^{\mu\nu}_{~~;\nu} 	
	= \varrho \, \rd_\tau U^\mu + g^{\mu \nu} p_{,\nu} \,
 ~~~~{\rm where}~~~~
	\rd_\tau \equiv U \cdot \nabla 
\label{eq:RelEoM} \eeq 
 stands for the generalization of the convective rate of change,
 i.e. the Lagrangian proper-time derivative.
 The projection of this equation along the four-velocity states energy 
 conservation while the perpendicular components express 
 conservation of momentum.

 	The specific Eulerian perturbation of the four-acceleration
 (normalized to the enthalpy) looks like 
 $ \udD \, U + \UdD \, u $, and one can use Eq [\ref{eq:RelDynSwi}] 
 to switch the dynamical variable in favor of the projected 
$\xi$'s\footnote{ 
 note that with this form of the stress-energy tensor, 
 $h^\alpha_{~\beta} \rd_\tau U^\beta \doteq \rd_\tau U^\alpha$,
 i.e. the four-acceleration automatically lies in proper 
 space-like hypersurfaces.
}:
\bea 	
	\hat{u} \! \cdot \!\! \nabla \, U 	+ 
	h \, (U \!\! \cdot \!\! \nabla \, \hat{u}^\alpha )
	&=& h (\UdD \, \dxi - \dxidD \, U) + 
	2 \, \dxidD \, U \cr 
&&\cr
 	&-& \gamma \xi^r \Omega_{,r} 
	\hat{\unvc}_\varphi \!\! \cdot \!\! \nabla \, U - 
	h \, \UdD ( \gamma \, \xi^r \, \Omega_{,r} \, \hat{\unvc}_\varphi)
\label{eq:Cumber} \eea
 where the silly hats on the $\xi$'s (signifying projected components)
 were dropped. 

 	The first term on the r.h.s. of Eq [\ref{eq:Cumber}]
 may be readily identified with 
 (the projection of) the Lie derivative of $\dxi$ along $U$.  
 Defining $q \equiv \unvc_t + \Omega \unvc_\varphi$ (so that $U = \gamma q$),
 one computes
\bea 		
	\Lie_U \dxi
	&=& \gamma \Lie_q \dxi - q \, \dxidD \gamma \cr
&&\cr 		
	&=& \ddot{\xi} 
	- \gamma \dxi^r \Omega_{,r} \hat{\unvc}_\varphi 
	+ \dxi \, \UdD \ln \gamma
	- U \, \dxidD \ln \gamma 
\eea
 where $\ddot{\xi} \equiv (i \sigma \gamma)^2 \xi$.
 Note that the last term disappears upon (re)projection 
 onto proper space-like hypersurfaces.

 	The second term on the r.h.s. of Eq [\ref{eq:Cumber}]
 is easily evaluated,
 $\dxidD \, U^\alpha = \gamma \dxi^r \Omega_{,r} \hat{\unvc}_\varphi^{_\alpha} 
 + \Gamma^\alpha_{\mu\nu} U^\mu \dxi^\nu$,
 and the third simply involves a projected affine connection.
 Evaluation of the (non-projected) last term yields four pieces:
\bea \UdD ( \gamma \xi^r \Omega_{,r} \hat{\unvc}_\varphi) 
 &=& \xi^r \Omega_{,r} \hat{\unvc}_\varphi \, \UdD \gamma 
 + \gamma \Omega_{,r} \hat{\unvc}_\varphi \, \UdD \xi^r 	\cr
&&\cr
 &+& \gamma \xi^r \hat{\unvc}_\varphi \, \UdD \Omega_{,r} 
 + \gamma \xi^r \Omega_{,r} \UdD \hat{\unvc}_\varphi.
\nonumber \eea
 Under the premise of negligible radial motion, 
 in the second piece above $\UdD \xi^r \simeq \dxi^r + \cO(U^r)$,  
 while the third is $\cO(U^r)$.  
 Likewise, pairing of all terms proportional to the logarithmic gradient 
 of the redshift factor yield same order (negligible) corrections 
 $(\dxi - \gamma \xi^r \Omega_{,r} \hat{\unvc}_\varphi) \UdD \ln \gamma
 \simeq \cO (U^r)$.
 Moreover, the last term above involves the same connection coefficient
 as the third term on the r.h.s. of Eq [\ref{eq:Cumber}].

	When all this is said and done, one gets for
 the specific Eulerian perturbation of the four-acceleration:
\bea
	\hat{u} \! \cdot \!\! \nabla \, U 	+ 
	h \, (U \!\! \cdot \!\! \nabla \, \hat{u})
	\stackrel{\cO(U^1)}{-\!\!\!-\!\!\!\longrightarrow}
	\ddot{\xi}^\nu 
	+ 2 \Gamma^\nu_{\alpha\beta} U^\alpha \dot{\xi}^\beta
	- 2 \Gamma^\nu_{\alpha\beta} U^\alpha (\hat{u} - \dxi)^\beta.
\label{eq:InerEoM} \eea
 The resemblance with Eq [\ref{eq:TheEssence}] 
 is remarkable but not accidental.

 	The shear(tidal) term is embodied by the third term on the r.h.s.:
 $\hat{u} - \dxi = - \gamma \xi^r \Omega_{,r} \hat{\unvc}_\varphi$.
 Note the non-trivially hatted unit vector 
 $\hat{\unvc}^\nu_\varphi = h^\nu_\mu \unvc^\mu_\varphi
	 = \unvc^\nu_\varphi + U^\nu U_\varphi$.
 We evaluate this term first using the standard form of the Kerr metric
 in the equatorial plane (Boyer-Lindquist coordinates):
\beq
ds^2 = -{\sD\over{\sA}} dt^2 + r^2\sA(d\varphi - \omega d t)^2
        + {1\over{\sD}}dr^2,
\label{eq:Metric} \eeq
 with $\omega \equiv {2 a / {\sA r^3}}$
 the rate of frame dragging by the hole 
 and where the metric functions of the radial {\bf BLF} coordinate are written
 as relativistic corrections (e.g. Novikov and Thorne 1973):
\[
\sA \equiv 1 + a^2/r^2 + 2 a^2/r^3,
~~{\rm and}~~
\sD \equiv 1 - 2/r + a^2/r^2,
\]
 in normalized geometrical units ($c = G = M_{_{\rm bh}} = 1$).

	In expanded form, the projection of the Killing vector associated with 
 the azimuthal symmetry is 
\[
 \hat{\unvc}_\varphi =
	[ 1 + \tilde{\gamma}^2 \tilde{r} v^{\tilde{\varphi}} \, \Omega ] 
	\unvc_\varphi 
    	+ \tilde{\gamma}^2 \tilde{r} v^{\tilde{\varphi}} \unvc_t, 
\] 
 where $\tilde{\gamma} = \gamma \sqrt{\sD/\sA}$ 
 is the redshift factor relative to ``locally non-rotating observers" 
 (Bardeen, Press \& Teukolsky 1972) and $\tilde{r} \equiv r \sA / \sqrt{\sD}$ 
 is the radius of gyration for the physical velocity in that frame,
 $v^{\tilde{\varphi}} = \tilde{r} (\Omega - \omega)$.

 Evaluation of the tidal term is a bit lengthy but straightforward 
\bea 	
	- 2 \Gamma^r_{\alpha\beta} U^\alpha (\hat{u} - \dxi)^\beta 
&=&	- {4 \over r^3} \, 
	\left\{\half \gamma^{2} \sD \, \rd_{\ln r} \Omega \right\}
	\left[ 
	\left( {\Omega \over {\Omega_+ \Omega_-}} - a \right)
	+
	\tilde{\gamma}^2 \tilde{r} v^{\tilde{\varphi}}
	\left( 1 - {\Omega \over {\Omega}}_+ \right) 
	\left( 1 - {\Omega \over {\Omega}}_- \right) 
	\right]
	\, \xi^r  
\label{RelTide} \eea
 where $\Omega_\pm = \pm (r^{3/2} \pm a)^{-1}$ refer to prograde 
 and retrograde circular orbits and 
 where the expression in the curly brackets equals 
 (minus) the shear of the congruence of circular, 
 equatorial geodesics (Novikov and Thorne 1973).
 
	Next, to evaluate the coriolis terms one finds $\xi^t$ from
 the gauge fixing condition $\xi \cdot U \doteq 0$.  Accordingly one finds
\bea
	2 \Gamma^r_{\alpha\beta} U^\alpha \dot{\xi}^\beta
	&=& ~~~~~2 \gamma \, {\sD \over r^2} 
	\left( {\Omega \over {\Omega_+ \Omega_-}} - a 
	+ (1 - a \Omega) \,
	\frac{r^2 \sA (\Omega - \omega)}{1 - {2 \over r}(1-a\Omega)} 
	\right) \, \dot{\xi}^\varphi, \cr	
&& \cr
	2 \Gamma^\varphi_{\alpha\beta} U^\alpha \dot{\xi}^\beta
	&=& - 2 \gamma \, {1 \over {r^4 \sD}} 
	\left( {\Omega \over {\Omega_+ \Omega_-}} - a + 
	2 r^2 \Omega 
	\right) \, \dot{\xi}^r \cr
{\rm and}~~~~~~~~&& \cr
	2 \Gamma^t_{\alpha\beta} U^\alpha \dot{\xi}^\beta
	&=& ~~ 2 \gamma \, {1 \over {r^4 \sD}} 
	\left( r^2 + a^2 - {a \Omega ( 3r^2 + a^2) 
} 
	\right) \, \dot{\xi}^r.	
\label{eq:RelCoriolis} \eea

	Eq [\ref{eq:InerEoM}] 
 for the (Eulerian) perturbation of the four-acceleration,
 $a^\mu = \hat{u} \! \cdot \! \nabla \, U^\mu + 
 h \, (U \! \cdot \! \nabla \, \hat{u}^\mu)$,
 was derived for the components of the Lagrangian displacement 
 in a coordinate frame that is fixed with respect to distant stars, 
 i.e. in the Boyer-Lindquist ``frame".
 However, because the instability is local 
 (at least for weak fields in thin disks)
 one needs to transform the components of Eq [\ref{eq:InerEoM}]
 for manipulation in terms of the local tetrad carried by co-moving observers. 
 This simply involves (matrix) multiplication by the basis vectors of 
 such a tetrad (e.g., Novikov \& Thorne 1973).  
 In our notation, the relevant basis vectors are 
 $e^{\hat{r}}_{\alpha} = 1/\sqrt{\sD} (0,1,0,0)$ and  
 $e^{\hat{\varphi}}_{\alpha} = \gamma r \sqrt{\sD} (-\Omega,0,0,1)$
 (note that transformation to the local tetrad yields equations for 
 non-coordinate motion, i.e. equivalent to motion in a local Cartesian basis).

 	Transformation of the $r$-component is trivial 
 (since one needs to transform both the acceleration and the displacement 
 vector in the basic EoM below, 
 the radial scale, $1/\sqrt{\sD}$, has no net effect):
\bea
	\sqrt{\sD} \, a^{\hat{r}}
 	&=& 	
	\ddot{\xi}^r 
	+ 2 \gamma \, {\sD \over r^2} 
	\left( {\Omega \over {\Omega_+ \Omega_-}} - a 
	+ (1 - a \Omega) \,
	\frac{r^2 \sA (\Omega - \omega)}{1 - {2 \over r}(1-a\Omega)} 
	\right) \, \dot{\xi}^\varphi \cr	
&& \cr
	&-& 
 	{4 \over r^3} \, 
	\left\{\half \gamma^{2} \sD \, \rd_{\ln r} \Omega \right\}
	\left[ 
	\left( {\Omega \over {\Omega_+ \Omega_-}} - a \right)
	+
	\tilde{\gamma}^2 \tilde{r} v^{\tilde{\varphi}}
	\left( 1 - {\Omega \over {\Omega}}_+ \right) 
	\left( 1 - {\Omega \over {\Omega}}_- \right) 
	\right]
	\, \xi^r. 
\label{eq:loc_r_acc} \eea
 On the other hand, 
 using the local azimuthal base vector, Eqs [\ref{eq:RelCoriolis}]
 and the gauge fixing condition $\xi \cdot U = 0$, 
 computation of the local $\varphi$-component, 
 $\propto -\Omega a^t + a^\varphi$, is a bit more involved
 (again, the radial scale factors out in the EoM and does not affect the 
 dispersion relation)
\beq
	{1 \over {\gamma r \sqrt{\sD}}} \, a^{\hat{\varphi}}
	= \left( \frac 
		{1 - {2 \over r} (1 - 2 a \Omega) - r^2 \sA \Omega^2}
		{1 - {2 \over r} (1 - a \Omega)} 
	 \right)
	\ddot{\xi}^\varphi
	- 2 \gamma \, {1 \over {r^4 \sD}} 
	\left( {\Omega \over {\Omega_+ \Omega_-}} - a 
	+ (1 - a\Omega) \Omega (3r^2 + a^2) 
	\right) \, \dot{\xi}^r.
\label{eq:loc_p_acc} \eeq

	Let us take a look at the Maxwell stress next.

	The fundamental premise of ideal MHD may be stated rather succinctly:
 in the rest frame of the fluid currents will flow uninhibited to 
 (instantaneously) cancel any hint of an electric field.
 A relativistic generalization of ideal MHD may be achieved by a similar
 covariant (albeit imperfect) postulate:
 $E_{\rm r.f.} = F \cdot U \doteq \emptyset$,
 e.g., the Faraday field tensor is ``purely magnetic" in the fluid frame. 

 	Such postulate brings a few mathematical consequences 
 (Phinney 1983):  \\
{\it i-} the second electromagnetic invariant vanishes everywhere, 
\beq
	\fourth F_{\mu\nu} \cF^{\mu\nu} \doteq \emptyset
\eeq
{\it ii-} the Faraday field tensor is Lie transported along 
 the worldlines of the fluid, 
\beq
	\Lie_U F \doteq \emptyset
\eeq
{\it iii-} the other zero eigenvector of the field tensor is the 
 (space-like) magnetic field $B \equiv \cF \cdot U$
 with $\cF^{\mu\nu} \equiv \epsilon^{\mu\nu\alpha\beta} F_{\alpha\beta}$ 
 the dual to the field tensor, 
\beq
 	F \cdot (\cF \cdot U ) \doteq \emptyset
\eeq
 and
{\it iv-} the field tensor is also invariant when transported along the
 magnetic field four-vector:
\beq
	\Lie_{_{\cF \cdot U}} F \doteq \emptyset
\eeq
	
 	Note further that, 
 since the four-velocity and the four-magnetic field are 
 orthogonal, $U \cdot B = 0$, 
 properties {\it ii and iv} above define a 2-surface of 
 invariance for the Faraday tensor
\beq
	\Lie_{_{aU + bB}} F \doteq \emptyset
\label{eq:RelIdealMHD}\eeq
 where {\it a} and {\it b} are arbitrary real numbers.
	
	Aside from the intrinsic (physical) difference in their spacetime 
 orientation, the mathematical similarities between $U$ and $B$ are 
 uncanny.  

	Let us go back to the Newtonian case for a moment. 
 From our definition of the Lagrangian variation of the three-velocity:
 $\rd_t \bxi \equiv \tilde{\Delta} \bV$, one finds the equation 
 governing the Lagrangian change of three-velocity: 
 $\tilde{\Delta} \bV = (\bV \!\cdot \!\Del) \, \bxi$,
 (e.g. Eq [\ref{eq:NewLag_v_Eul}]). 
 As noted in the footnote,
 the difference between $\tilde{\Delta}$ and $\Delta$ 
 is related to the choice of gauge for $\xi^\alpha$.
 The induction equation of non-relativistic MHD yields a 
 virtually identical relation for the magnetic field variation
 (in a frame where the fluid was originally at rest) 
 which is spoiled by fluid compressibility:
 $\tilde{\Delta} \bB 	= (\bB \!\cdot \!\Del) \, \bxi 
			- \bB \, \Del \!\cdot \!\bxi$.
 Nevertheless, making use of the continuity equation and  
 weighting the field by the inverse of the 
 fluid's mass density 
 $\tilde{\bB} \equiv \bB / \rho$
 cleans up its connection to the displacement vector field
 $\tilde{\Delta} \tilde{\bB} = \tilde{\bB} \!\cdot \!\!\Del \, \bxi$. 
 If only conservative forces, $U \cdot f \doteq 0$, act on the fluid,
 it can be shown that use of energy conservation in lieu of mass conservation
 simply swaps the fluid's rest mass density by the relativistic enthalpy
 (a world scalar), above.
 Thus, we choose to work below with a specific measure 
 of the magnetic four-vector weighted by the inverse of the 
 fluid's relativistic enthalpy $\tB \equiv \inventh (\cF \!\cdot \!U)$.
 Such combination of observables (and its perturbation) 
 occurs naturally in the problem at hand.

   	Applying the Lagrangian variational operator,
 c.f. Eq [\ref{eq:RelLag_v_Eul}], on $\tB$ under 
 the constraints from ideal MHD noted above, Eq [\ref{eq:RelIdealMHD}],
 yields (contrast this with Eq[\ref{eq:U_Var}])
\beq	\Delta \tB \doteq \emptyset = \delta \tB + \Lie_\xi \tB. 
\label{eq:EulVarB} \eeq
 This equation states a manifestly covariant expression for the 
 Eulerian perturbation of the (enthalpy-weighted) Faraday tensor 
 under ideal MHD constraints: $\tilde{b} = \Lie_{\tB} \xi$.
 Demanding that the total magnetic field four-vector 
 be orthogonal to the (unperturbed) four-velocity is equivalent to 
 projecting its Eulerian perturbation into proper spacelike hypersurfaces: 
 $\tilde{b}^\mu \rightarrow \hat{b}^\mu = h^\mu_\nu \, \Lie_{\tB} \xi^\nu$.
 Again, we suppress the hats below while tacitly imposing the condition
 $\xi \cdot U \doteq \emptyset$ throughout.
 
 	The Eulerian perturbation of 
 the specific measure of the Lorentz force, 
 $\delta (\tilde{T}_{_{\rm e.m.}})_{~;\nu}^{\mu\nu} = 
 \delta F^{\mu\nu} \tilde{J}_\nu + F^{\mu\nu} \delta \tilde{J}_\nu$,
 may now be written in terms of the Lagrangian displacement
 but the general expressions are not particularly illuminating.
 Evaluated in the frame where the fluid was originally at rest,
 the Eulerian perturbations of the field tensor and 
 of the four-current depend linearly on the components of $b$: 
 $\delta F = F (b)$ and 
 $4\pi \, \delta \tilde{J} =  \rd \cdot \delta \tilde{F} (b)$.

 	We proceed by assuming negligible gradients 
 of the background specific field (${\nabla \tB \doteq \emptyset}$): 
\beq 	\tilde{b} 	= \tB \!\cdot \!\nabla \xi 
			- \xi \!\cdot \!\nabla \tB 
		{\longrightarrow}  \tB \! \cdot \! (i k) \, \xi,
\label{eq:Simple} \eeq
 and, consistent with this assumption, we also ignore the 
 $\delta F \cdot \tilde{J}$ term in the perturbation of the Maxwell stress
 (i.e. gradients of the background field tensor ($\propto J$)
 gentler than those of the perturbations).


	The simple ``linear poking" of the field 
 tensor may now be written in a manifestly covariant manner 
\beq
 	F^{\mu\nu} \delta \tilde{J}_\nu 
	= {1 \over \varrho} (B \cdot ik)^2 \, \xi^\mu.
\label{eq:spring} \eeq  
 Naturally, evaluation of the 
 elastic response of the field is straightforward in the rest 
 frame of the fluid where one has
 $F^{\mu\nu} \delta \tilde{J}_\nu 
	\doteq - (\Valf k_{\hat{\bB}})^2 \, \xi^\mu$.	
 The only difference with the non-relativistic analog is that 
 the Alfv\'en speed is now weighted by the relativistic enthalpy of the fluid
 $\varrho \Valf^2 \equiv \half F^{\mu\nu} F_{\mu\nu}$.

 	We are now all geared up to put together the pieces of the puzzle. 
  	In terms of the Lagrangian displacement vector field,
 the r.h.s.'s of Eqs [\ref{eq:loc_r_acc} \& \ref{eq:loc_p_acc}]
 are to be balanced by the elastic response of the field tensor
 to the poking by $\xi$, c.f. Eq [\ref{eq:spring}]
 (note that the radial scales of the transformation 
 into the rest frame of the fluid cancel one another).
 This balance is locally equivalent to 
 $a^\mu = - q_{\hat{\bB}}^2 \xi^\mu$, i.e.
 the covariant components of the fluid's acceleration respond to 
 a force proportional to the displacement vector 
 (with the unnormalized ``spring constant" 
 $q_{\hat{\bB}} = (\Valf k_{\hat{\bB}})$ provided by the field).  
 By construction, both of these vectors are orthogonal to $U$ and collinear.
 Furthermore, since $\ddot{\xi}^\mu \equiv (i\gamma\sigma)^2 \xi^\mu$ and
 $\gamma \sigma$ is a world scalar 
 to be identified with the true co-moving frequency 
 (as measured by an observer riding along with the fluid),
 it follows that 
 $\ddot{\xi}^{\hat{\varphi}} \equiv (i\gamma\sigma)^2 \xi^{\hat{\varphi}}$.
 With these relations and the aforementioned equations
 for the tidal and coriolis terms,
 one arrives to lengthy component equations for $\xi^r$ and $\xi^\varphi$,
 for general $\Omega \equiv U^\varphi/U^t$ and negligible radial flow. 

 	In the case of circular geodesic flow, the equations simplify
 beautifully (horizontal regime) 
\bea
	\ddot{\xi}^r  
	- 2 \gamma \, {\sD \over r^{1/2}} \, \Omega_{\pm}
	\left(  	
	\frac	{r^3 - 3 r^2 \pm 2ar^{3/2} }
		{ {r^{3/2} \pm a} - 2r^{1/2}} 
	\right) \, \dot{\xi}^\varphi 
	- {4 \over r^{3/2}} \,  
	\left\{ \3ov4 \gamma^{2} \sD \, r^{3/2}\Omega^2_\pm \right\} 
	\, \xi^r
&=& 	- q_{\hat{\bB}}^2  \, \xi^r 	\cr
&&\cr
	\ddot{\xi}^\varphi 
	+ 2 \gamma \, {1 \over {r^{5/2} \sD}} \, \Omega_\pm
	\left( {r^{3/2} \pm a - 2r^{1/2}} \right)
	 \, \dot{\xi}^r
~~~~~~~~~~~~~~~~~~~~~~~~~~~~~~~~~~
&=& 	- q_{\hat{\bB}}^2  \, \xi^\varphi.	
\label{eq:Hor_EoM} \eea
 These immediately yield the sought after 
 dispersion relation near a rotating hole
\beq
	(\gamma\sigma)^4
	- 
	\left[ 
	{4 \gamma^2} \Omega_\pm^2 
	\left( \sC_\pm - \3ov4 \sD \right) + 2  q_{\hat{\bB}}^2 
	\right] (\gamma\sigma)^2 
	+
	q_{\hat{\bB}}^2 \,
	\left[ q_{\hat{\bB}}^2 - {4} \, 
 	\left\{\3ov4 \gamma^2 \sD \, \Omega^2_\pm \right\} \right]
	= \emptyset
\eeq
 where $\sC_\pm \equiv {1 - {3\over r} \pm { 2a \over r^{3/2}} }$ 
 corresponds to the $\sC$ function of Novikov \& Thorne (1973)
 for prograde orbits.

	Factoring out the 
extrinsic\footnote{
 As defined, $\Omega \equiv U^\varphi / U^t$ reflects motion as observed 
 in the Boyer-Lindquist frame, i.e., in a frame extrinsic to the 
 fluid.  It follows that the timescale associated with $\Omega^{-1}$  
 does not reflect a proper dynamical timescale.
} 
 dynamical frequency,
 $\Omega_\pm$, one arrives to the normalized dispersion relation
 (with $\gamma \sigma \equiv \Omega_\pm \, \hat{\sigma}$)
\beq
	\hat{\sigma}^4 
	- 
	\left[ \qB^2 + \hchi_\pm^2 
	\right] \hat{\sigma}^2 
	+
	\qB^2 \left[ \qB^2 +  4 \AM \right]
	= \emptyset
\label{eq:GenRelDisp} \eeq
 where 
\[ \AM \equiv - \left\{\3ov4 \gamma^2 \sD \right\}
~~{\rm and}~~
\hchi^2_\pm = 4 \gamma^2 \left( \sC_\pm - \3ov4 \sD \right)
\]
 denote the normalized shear parameter and (co-moving) epicycle frequency 
 (note that ${1 \over \gamma} \, \hchi$ corresponds to the well known 
 result of epicycle frequency as measured at asymptotic infinity).  
 One thus sees that 
 with the proper generalizations of the epicycle frequency and shear 
 parameter, the local dispersion relation is identical with the Newtonian
 case in the limit of no fluid compression
 and $\unvc_{\hat{\bB}} \cdot \bxi \doteq \emptyset$ 
 (i.e., the ``classical" Balbus-Hawley instability, Eq[\ref{eq:disper}]). 

	Using the relation $\gamma^2 = (1 \pm a/r^{3/2})^2 \, \sC_\pm^{-1}$
 for cold, circular, {\it geodesic} flow (Novikov and Thorne 1973), 
 one finds the fastest growing modes to conform with
\beq
	\qB^2 	= 1 - \inv16 \hchi^4 
		= 1 - 
		\left( 1 \pm {a \over {r^{3\over2}}} \right)^4 \,
		\left\{ 1 - {3\over4} \, {\sD \over \sC_\pm} \right\}^2
\label{eq:WaveNumb} \eeq 
 which remains finite and close to the Newtonian value of ${15 \over 16}$
 for all radii {\it outside} the ISCO
 (and for any value of the rotation parameter). 
	
 	To attach meaning to the polynomial functions that appear
 naturally in the dispersion relation for the magnetorotational instability,
 recall the range of radii that define particle dynamics in the Kerr geometry
 (Bardeen \etal 1972):  
\begin{enumerate} 
	\item The marginally stable circular orbit (a.k.a. the ISCO), 
 	 $r_{\rm ms}$, 
	 corresponds to the root of $\hchi_\pm = 0$. 
 	\item The radius of the circular photon orbit, $r_{\rm ph}$,
	 is where $\sC_\pm = 0$.
	\item The event horizon, $r_+$, happens at the outer root of $\sD = 0$. 
\end{enumerate} 

	One therefore has the following ordering of radii 
 for any value of the rotation parameter 
 $a$: $r_{\rm ms} > r_{\rm ph} > r_+$.
 As remarked by Bardeen \etal '72, 
 when $a = 1$, the proper radial distance between these 
 radii is non-zero in spite of ``coinciding with the horizon",  
 i.e. in spite of laying at the same Boyer-Lindquist radial coordinate. 
 
	Inspection of Eq [\ref{eq:WaveNumb}] now shows that
 $\qB \rightarrow 0^+$ as $r \rightarrow r^+_{\rm ph}$ so the most unstable 
 MRI modes go to large scale just outside the photon orbit.  
 Moreover, utilizing that expression for $\qB$ 
 in the unstable root of the dispersion relation, 
 one finds the growth rate (or frequency!) to be given by 
\beq
	-\hat{\sigma}^2 = \left\{ {3\over4} {\sD \over \sC_\pm} \right\}^2 
		\left[
			\left(1 \pm {a \over r^{3\over2}} \right)^4 
			- 
			{8\over3} {\sC_\pm \over { \sD}} 
			\left( 	\pm 2 {a \over r^{3\over2}}  
				+   5 {a^2 \over r^{3}}
				\pm 4 {a^3 \over r^{9\over2}}
				+     {a^4 \over r^{6}}
			\right) 
			+
			\, \left({4 \over 3} {\sC_\pm \over {\sD}}\right)^2
			\left(	    4 {a^2 \over r^{3}}
				\pm 4 {a^3 \over r^{9\over2}}
				+     {a^4 \over r^{6}}
			\right) 
		\right].
\label{eq:Growth} \eeq

	For a non-rotating hole, 
\[ 	\qB \rightarrow 0 ~~@~~r = r_{\rm ph} (1 + 
	{\fifth} ), 
	~~~{\rm and~ the~ local~ growth~ rate}~~ 
	\hat{\sigma} = {3\over4} {\sD \over \sC_\pm} 
		 \rightarrow 2
\]
 while for a rotating hole,
 the MRI quenching radii (for fastest growing modes) also occur 
 just outside the circular photon orbit 
 and may be readily extracted from the above relations.
 In Figs 1 and 2, we plot the general relativistic modifications 
 the fastest growing linear wavemodes, 
 wavenumbers and growth rates respectively, 
 as functions of radius and for different values of spin parameter $a$.  
  
	To go beyond this point, one would need to address global effects 
 arising, for instance, from field curvature terms 
 (see, e.g., Curry \& Pudritz 1995, Ogilvie \& Pringle 1996)
 and from the non-negligible radial velocity profile.
 Further investigation of the nature of the global 
 instability is beyond the scope of this paper.

\section{Discussion}

	The 
 MRI--in its most simple, local, incompressible variant--is
 found to operate virtually unabated down to the marginally stable orbit
 for massive particles.  This radius is nearly coincident with the putative
 inner boundary of standard, thin accretion disks in the Kerr geometry. 
 A vanishing epicycle frequency at $r_{\rm ms}$ means that 
 the fastest growing wavenumbers tend to be of a bit smaller scale, 
 $\qB^2: \15ov16 \rightarrow 1 - \cO ({a r^{-3/2}})$,
 while growing faster than classically, 
 $i\hat{\sigma}:  \3ov4 \rightarrow 1 + \cO ({a r^{-3/2}})$.
 The effects of strong gravity become truly significant only in a regime
 where circular, cold, geodesic flow is unstable
 (i.e., where $\hchi^2_\pm < 0$). 

	Recall that particle trajectories with $U^\varphi/U^t = \Omega_\pm$
 exist inside $r_{\rm ms}$ and all the way down to $r_{\rm ph}$ but,
 in the presence of turbulent velocity fluctuations, 
 body forces such as a radial pressure gradient would be required 
 to confine the flow to such circular orbits.  
 Although very little is concretely known about the accretion flow
 inside $r_{\rm ms}$, two rather robust remarks may be ascertained:
 The flow inside $r_{\rm ms}$ cannot be supported 
 centrifugally and it must therefore deviate from a standard thin disk.
 In addition, depending on the timescale for infall, 
 the flow may not have time to cool significantly and advection of entropy 
 will become progressively more important as $r_+$ is approached.
 {\sl A robust prediction of this paper is the expectation that
 free energy tapping from the differential shear flow goes on in the 
 region immediately below $r_{\rm ms}$}.
 
 	One may envisage the situation inside the ISCO to evolve from a 
 Mildly Advective Accretion Flow (MAAF) to a fully Advection Dominated 
 Accretion Flow (ADAF) as the photon orbit is approached.  
 In fully or partly advective accretion flows, 
 such as those modeled by Popham \& Gammie (1998),
 the angular velocity profile ``peaks" precisely at $r_{\rm ph}$ and 
 quickly drops therein to match the angular velocity at $r_+$.
 More importantly, when cooling by advection of entropy is moderately
 important--say, for advection fractions $f \simeq$ a few percent--the
 angular velocity profile departs very slowly from circular 
 geodesic flow, $U^\varphi/U^t \simeq \Omega_\pm$
 down to a region below the marginally bound orbit.
 The transition from nearly Keplerian to plunging orbits can be clearly seen 
 in one of the very few global slim disk models where the cooling fraction
 is calculated explicitly: the 1D models of Popham \etal (1998,
 albeit in the exotic scenario of a hyper-accreting black hole). 
 In these models the radial velocity component is non-negligible when
 compared with the local sound speed 
 (the sonic point generally occurs below $r_{\rm ms}$, 
 even near $r_{\rm mb}$ for low values of $\alpha$),
 but $v^{\tilde{r}}$ is generally smaller than $v^{\tilde{\varphi}}$ 
 down to the region below $r_{\rm mb}$.
 (Note that the radial speed in the corotating frame, 
 e.g. Gammie \& Popham's (1998) $V$, 
 is related to the same in the locally non-rotating frame
 by $v^{\tilde{r}} = \gamma_{\varphi}^{-1} V$.)
 
 	The major limitation of the work presented herein is the presumption 
 of negligible radial flow which greatly simplifies matters from the onset 
(see Eq [\ref{eq:Eul2Lag_D}]).  
 At this point, it is unclear how much the results will change 
 when full consideration is made for the radial inflow.
 Since the changes could be qualitatively 
 significant--recent reports negate the reversal of the centrifugal force 
 when the radial speed overwhelms the azimuthal component 
 (Mukhopadhyay \& Prasanna 2001, Prasanna 2001)--this
 point should be the subject of close scrutiny in a future paper.
 Meanwhile, the adoption of an angular velocity profile corresponding
 to circular equatorial geodesic orbits seems a reasonable 
 rough approximation in view of the above observations of advective flows.
 In this spirit, we argue below that the natural evolution of the MRI
 inside the marginally stable orbit is at least consistent with 
 this assumption.

\begin{figure}
\hbox{~}
\centerline{\psfig{file=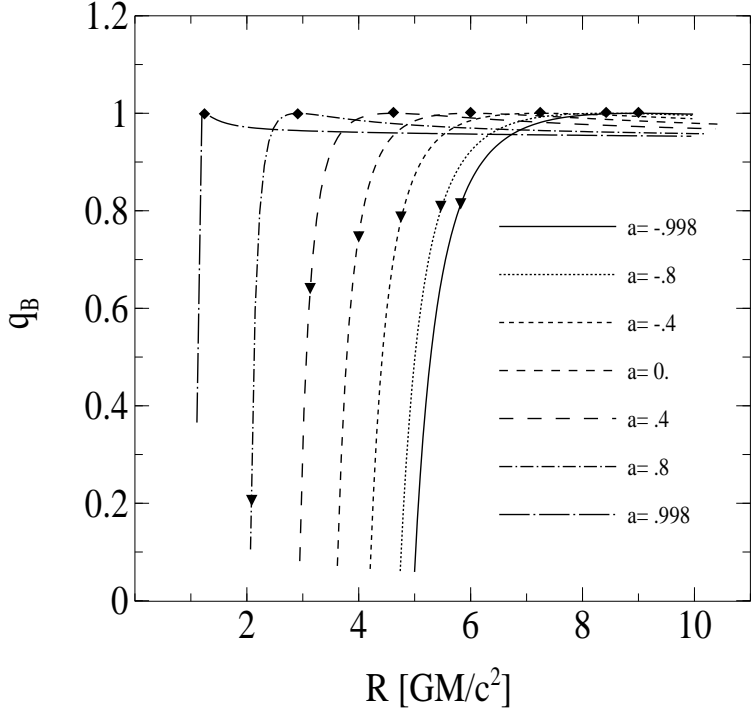,angle=0,width=5truein}}
\vskip 1mm 
\caption{Normalized wavenumber, $\qB$, as a function of radius
(in gravitational radii)
for several values of the spin parameter $a$.
Diamonds indicate the location of the marginally stable orbit,
$\hchi \doteq \emptyset$,
and triangles, the location of the marginally bound orbit.
}
\end{figure}

\begin{figure}
\hbox{~}
\centerline{\psfig{file=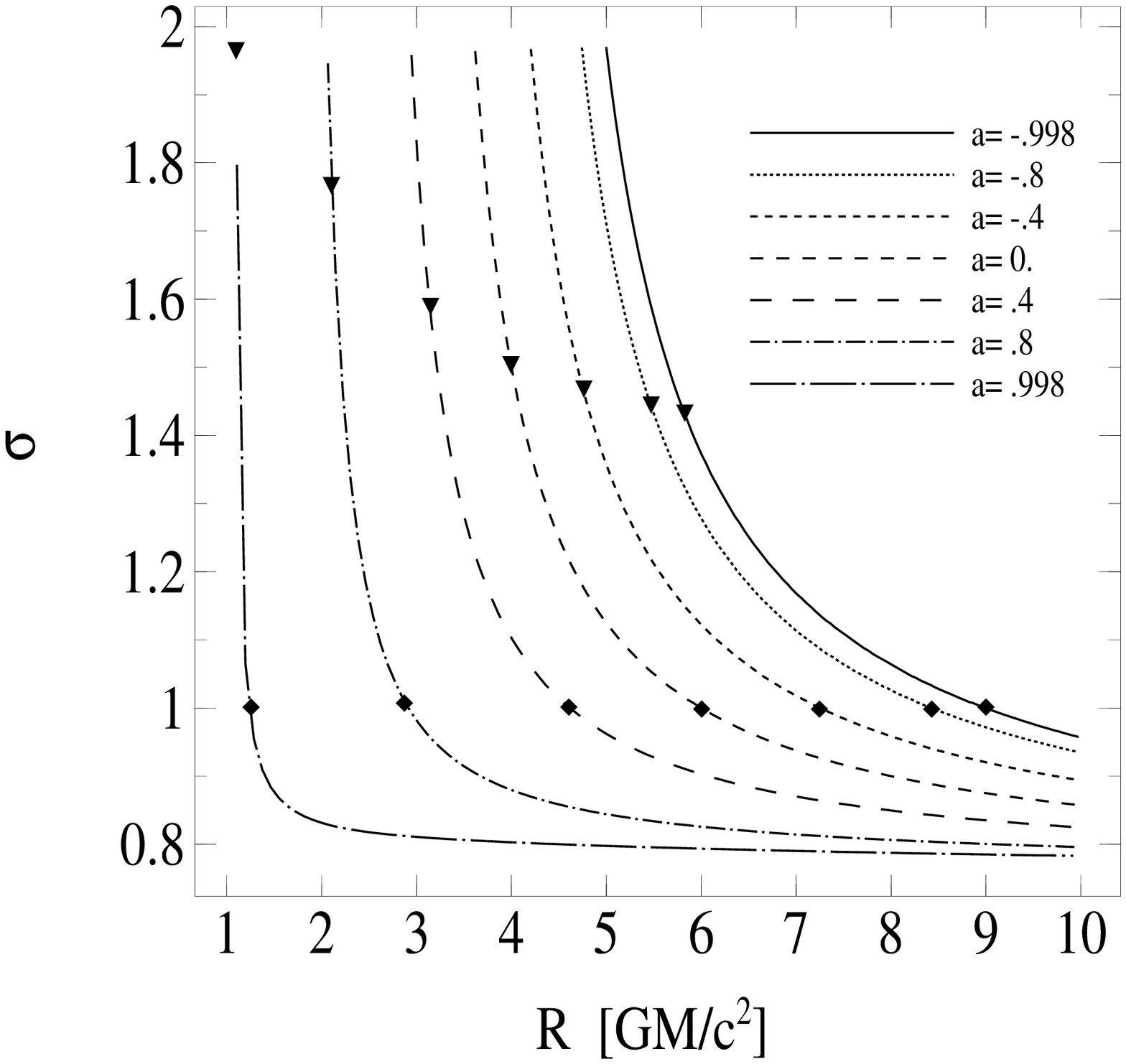,angle=0,width=5truein}}
\vskip 1mm 
\caption{Normalized growth rate, $\hat{\sigma}$, as a function of radius
for several values of the spin parameter $a$.
}
\end{figure}

	Assume, in quasi-linear fashion, that the time- and 
 length-scales provided by the linear dispersion relation reflect 
 the growth and size of the dominant turbulent eddies to within factors 
 of order unity to a few.  
	Provided that $v^{\tilde{r}} \ltaprx~ v^{\tilde{\varphi}}$,
 simple linear growth/non-linear decay arguments 
 (e.g., \AG 1999) can be used to 
 predict a predominantly toroidal field topology: 
 The MRI constantly promotes radial/azimuthal field growth 
 from ``horizontal" velocity fluctuations, 
 $\xi^{\hat{r}} \simeq -\xi^{\hat{\varphi}}$, 
 while the coherent, background azimuthal shear flow converts this field
 into toroidal field at twice the rate of radial field generation.
 In the rest frame of the fluid, 
 the tapping of free-energy associated with the shear flow 
 becomes very rapid as the flow turns relativistic.  
 Indeed, a co-moving observer measures the shear parameter, 
 $2 {\rm A} / \Omega_\pm$ to be ${3\over 2} \gamma^2 \sD 
 \simeq {3\over 2} \, (1 \pm a/r^{3/2})^2 \, \sD/\sC_\pm$,
 higher than the ``Keplerian" frequency 
 associated with the global dynamical timescale as seen at large distances 
 (the redshift factor comes in because we chose to measure 
 the angular frequency in terms of Boyer-Lindquist coordinates).
 
	The ratio $\sD / \sC_\pm$ represents a gauge of the relative strength 
 of two inertial terms, shear and coriolis.  
 Setting aside the issue of radial flow for a moment, 
 our dispersion relation suggests
 that as material approaches the region just outside of the photon orbit
 where $\sC_\pm$ vanishes, the slow branch of the dispersion relation 
 (i.e. the MRI) is stabilized by the predominance of shear 
 over the coriolis terms.
 Recall that the location of the circular photon orbit is 
 the place where the centrifugal force reverses its direction: 
 Inside $r_{\rm ph}$,
 increasing the velocity of a test particle pulls it in further
 (see, e.g., Abramowicz \& Prasanna 1990 and references therein).
 The limit of $\qB \rightarrow 0^+$ means that what was 
 essentially a local instability becomes a global phenomenon. 
 Although such a regime is formally outside the scope of the local analysis,
 one can anticipate a few rather interesting qualitative consequences.

 	At first glance, the dispersion relation Eq [\ref{eq:GenRelDisp}] 
 shows the appearance of an interchange, radially buoyant mode 
 (T. Foglizzo priv. comm., \AG 1999).
 More likely, this would simply imply the need for a steep radial 
 stratification profile.  Indeed, 
 if the coherence lengthscale of the field were to reach 
 the comoving length associated with the radial scale, $\sA r/\sqrt{\sD}$,  
 the disk could make a 
 transition from centrifugally driven to magnetically driven: 
 {\sl MRI modulated dynamics guarantee that the \Alf speed associated with the
 toroidal field at this large scale would be comparable to the orbital speed.}
 Moreover, the field generated at large scales is less prompt to decay through
 reconnection and also more buoyant.
 This has very important consequences for the energy fraction going 
 into--and persisting in--electromagnetic channels. 

	The radial velocity profile will very likely change 
 the expected outcome once the radial velocity becomes 
 supersonic or super-Alfvenic,
 but some of the qualitative features of the this analysis may carry over
 when the full problem is solved, analytically or otherwise.
 If so,
 in this part of the so-called ``plunging region" of the flow, 
 the turbulent eddies will tend to grow larger
 while the field direction will tend to track the surfaces of null angular
 velocity gradients (no longer purely toroidal). 
 The implied field topology is that of large-scale horizontal field domains.

\subsection{effects of radiation stress and neutrino trapping}

	Precise assessment of the dynamical role of radiation in the general
 relativistic regime is hampered by the breakdown of one key assumption 
 made to simplify the ``linear poking" on the Faraday field tensor:
 Use of the enthalpy weighted specific four magnetic field 
 in Eq [\ref{eq:spring}]. 
 On the other hand, one expects
 a photon gas--semi-contained by a neutral plasma through Compton 
 scattering--to comprise a rather funny MHD fluid where the
 magnetic field is truly frozen only to the co-moving volume 
 associated with the mass density but for which
 pressure perturbations do not behave adiabatically.
 It follows that 
 when the fluid transitions into radiation pressure domination, 
 compressive modes (e.g. toroidal field, non-axisymmetric modes) 
 may lose pressure support in an unfavorable range
 of wavenumber phase-space (Agol \& Krolik 1998). 
 One can prove that the MRI falls squarely
 in such radiative heat conduction damping regime 
 (Blaes \& Socrates 2001).
 \AG \& Vishniac (2002) show that 
 the behavior of the energy equation is in some (algebraic) sense 
 ``quasi-adiabatic" for exponentially growing, non-propagating modes.
 Mathematically, this means that a real, analytical, 
 slow-varying function of the scale of the perturbations, 
 $\tilde{\Gamma} (i \tilde{\bk}^2/\tilde{k}^0)$,  
 can be used to treat the energy equation in quasi-adiabatic fashion. 
 Radiative heat conduction isotropizes the modes and, 
 to zeroth order, 
 one can use such quasi-adiabatic index in Eq [\ref{eq:qB-om}]
 to anticipate that the effects of radiative heat conduction out of
 compressive toroidal modes is to increment the threshold of 
 shear parameter where $\qB \rightarrow 0^+$ 
 from $-2$ to $-\AM \rightarrow 1 + 2 \sD / (1 + \Lambda)$.
 Nevertheless, since $\AM \propto \sD/\sC_\pm$ and 
 $\sC_\pm \rightarrow 0$ @ $r_{\rm ph}$, 
 the increase in shear threshold in this setting is rather inconsequential.

	Note further that
 	the qualitative nature of energy deposition in radiation pressure
 dominated fluids is insensitive to the details of the (global) cooling 
 but it is explicitly sensitive to the optical thickness of the relevant
 eddies.
 Thus, upon the onset of neutrino trapping in the neutrino cooling regime
 of hyper-accreting black holes,
 one may reasonably expect MRI modulated dynamics at $p_\nu \gtaprx \Prg$
 (gas and radiation are tightly coupled)
 to resemble the standard disk case when $\Prad \gtaprx \Pgas$.
 Turner \etal 2001 report that 
 the non-linear outcome of the MRI in this setting is a porous medium
 with drastic density contrasts
 as to cheat the Eddington limit at high accretion rates.
 Under nearly constant total pressure and temperature, 
 the non-linear regime shows that density enhancements 
 anti-correlate with azimuthal field domains 
 (just as expected from the linear theory)
 and that turbulent eddies live for about a dynamical timescale
 while mass clumps are destroyed
 through collisions or by running through a localized region of shear.

	Since the turbulent eddies in the disk are largely 
 instabilities of the toroidal field (at moderate values of the field), 
 large-scale horizontal field domains near the marginally bound orbit 
 would naturally force the baryonic component of the accretion flow into
 spatially segregated, massive clumps that occur near the nodes of 
 non-axisymmetric (toroidal) MRI eddies (\AG \& Vishniac 2002). 
 This expectation motivates the picture of massive clumpy accretion 
 suggested in the introduction.  

\section{Ending Notes}
\label{sec:EndNot}

	In summary, this work shows that the MRI is virtually unaffected
 by strong gravity outside the innermost stable circular orbit. 
 Secondly, it indicates that the instability becomes non-local inside 
 this region. 
 Indeed, the MRI may leave behind a large scale, ordered field 
 as the fluid heads in towards the circular photon orbit
 (with an orientation that tracks surfaces of null angular velocity gradients).
 Assuming incompressibility and the angular velocity profile 
 of circular geodesic flow, the fastest growing modes die off while 
 going to large scales at a radius just inside the marginally bound
 orbit.  
 Accountability of compressibility as required to address the effects
 of radiation stress will bring the critical MRI quenching radius in, 
 slightly closer to the photon orbit.
 Radiation stress, when significant, 
 will diminish the growth rate while increasing the threshold 
 of shear parameter to quench the MRI.

	Radial inflow will affect the global field topology but the details
 depend on poorly understood fluid trajectories in a region where cold, 
 circular geodesic flow is unstable.
 As was pointed out by Krolik (1999), the standard assumption of ballistic
 orbits is never self-consistent for ideal MHD accretion inside $r_{\rm ms}$. 
 Indeed, when magnetic turbulence is the culprit of angular momentum 
 transport in the disk, the magnetic field energy density must become 
 comparable to the rest-mass energy density of the fluid in the plunging 
 region.  Yet, unlike Krolik's suggestion, 
 we do not believe that {\it linear} 
 \Alf waves could efficiently transport energy 
 from inside $r_{\rm ms}$; the magnetic field there is still highly unstable,
 and the range of stability of such waves is limited by inertial forces. 

 	On the other hand, in this paper we demonstrate that  
 energy deposition and angular momentum transport through the MRI 
 go on virtually unscathed in the region just below $r_{\rm ms}$.  
 An important note is the promptitude of this process at near Eddington rates
 since the MRI directly feeds the photon bath through compressive damping
 of the modes (\AG \& Vishniac 2002).  
 Energy deposition into the radiation field thus occurs on the MRI timescale!
 On the other hand, near $r_{\rm ph}$ 
 the flow will inevitably transition into advective cooling.
 Assessing the magnetic field dynamics in the region 
 $r_{\rm ms} > r \gtaprx \, r_{\rm mb}$
 is essential to predict the efficiency of accretion, 
 and to address some large scale effects such as jet launching 
 and disk-hole coupling.

 	At highly super-Eddington accretion rates
 (such as those expected in the prompt stages of hyper-accreting 
 black hole formation), the fluid may possess
 non-trivial amounts of internal energy per unit rest mass of baryons. 
 For such a hot MHD fluid, $r_{\rm ms}$ does not represent a significant 
 boundary to the disk/flow and such may occur rather closer to 
 $r_{\rm ph}$. 
 This stresses the importance of addressing MHD processes 
 in the region above the circular photon orbit. 
 Along these lines, 
 we have motivated the provocative conjecture that copious 
 gravitational wave losses ensue through black hole ringing when a 
 hyper-accreting black hole enters the accretion regime where 
 neutrino trapping occurs.  
 This argument, which combines linear regime phenomenology 
 with the latest numerical results from accretion in 
 radiation-stress dominated environs, 
 leads to a picture of near-hole accretion where large-scale 
 horizontal field domains channel the flow into massive clumps that 
 ``thump" the hole. 

 	Lastly, note that
 a strong, toroidal field topology is ripe ground for MHD 
 instabilities that promote poloidal field generation such as the 
 Parker and radial interchange 
 instabilities (in the vertical and horizontal regime respectively). 
 This instabilities could provide a physical justification for desirable 
 field topologies invoked in jet launching 
 and the Blandford-Znajek processes.

\section*{Acknowledgments}

  	It is a great pleasure to acknowledge stimulating and instructive 
 discussions with Lee Lindblom, Roger Blandford, Ethan Vishniac, 
 Thierry Foglizzo and Sterl Phinney.
 I am also very thankful to an anonymous referee for useful comments 
 which helped improve the manuscript.  
 I am greatly indebted with Caltech's 
 Theoretical Astrophysics and Relativity Group for their hospitality.

\pagebreak 

{\footnotesize

\end{document}